\documentclass[letter]{aa}

\usepackage{graphicx}
\usepackage{natbib}
\bibpunct{(}{)}{;}{a}{}{,}  

\usepackage{txfonts}

\def \f#1{\ensuremath{\vec{#1}}}
\newcommand{\Om}{{\it \Omega}}
\begin{document}

\title{Helicity generation and $\alpha$-effect by Tayler instability with $z$-dependent differential rotation}

\author{M. Gellert \and G. R\"udiger \and D. Elstner}

\institute{Astrophysikalisches Institut Potsdam, An der Sternwarte 16, D-14482 Potsdam\\
           \email{mgellert@aip.de, gruediger@aip.de, delstner@aip.de}}

\date{\today}

\abstract
 {}
 { 
  We investigate the instability of toroidal magnetic fields resulting from the action of $z$-dependent differential rotation 
  on a given axial field $\vec B^0$ in a cylindrical enclosure where, in particular, the helicity of the resulting nonaxisymmetric 
  flow  is of interest. We probe the idea that helicity is related to the external field and the differential rotation 
  as ${\cal H} \propto B^0_i\, B^0_j\, \Om_{i,j}$. 
   }
  {
  We conduct isothermal magnetohydrodynamic simulations of a quasi-incompressible medium with finite viscosity and
  conductivity in a perfectly conducting container, and analyze both the kinematic and current helicity of the resulting
  field by regarding the nonaxisymmetric parts of the field as fluctuations.
  }
  { 
  The observed instability leads to a nonaxisymmetric solution with dominating mode $m=1$. With the onset of instability, 
  both kinematic and current helicity are produced which fulfill the suggested relation ${\cal H} \propto B^0_i\, B^0_j\, \Om_{i,j}$.  
  Obviously, differential rotation $\mathrm{d}\Om/\mathrm{d}z$ only needs an axial field $B^0_z$ to produce significant helicity. 
  Any regular time-dependency of the helicity could not be found. The resulting axial $\alpha$-effect $\alpha_{zz}$ is mainly due to the 
  current helicity, the characteristic time scale between both the values is of the order of the rotation time. If the axial field is 
  switched off, then the helicity and the $\alpha$-effect disappear, and a dynamo is not observed.
  } {}

\maketitle

\section{Introduction}

Helicity in rotating turbulence plays a basic role in dynamo theory for the generation of large-scale cosmic magnetic fields \cite{krause_1980,brandenb_2005}.
Usually stratification of density or turbulence itself is needed to generate helicity. We demonstrate that, even without stratification, helicity can exist 
due to the common action of magnetic fields and differential rotation. Whereas differential rotation depending on latitude is well-known from observations 
of star spots, differential rotation (of the interior of stars) depending on the $z$-coordinate only, appears more rarely. It is predicted and observable, 
for instance, in the tachocline of the sun (R\"udiger \& Kitchatinov 1997; Kitchatinov \& R\"udiger 2005), and influences the internal rotation of massive 
stars \cite{maeder}. It may open the possibility of studying nonaxisymmetric (large-scale) structures by shear-driven magnetohydrodynamic (MHD) instability 
of laminar flows \cite{braithspruit_2004,braithw_2006}.\\ 
Differential rotation transforms poloidal field components into toroidal components, which due to the Tayler instability (TI), become unstable if a critical 
amplitude is exceeded (Vandakurov 1972). Tayler (1973) showed that a magnetic field $B_{\phi}$ becomes unstable against nonaxisymmetric perturbations if 
the condition
\begin{equation}
\frac{\mathrm{d}}{\mathrm{d} R} \left(R B_{\phi}^2\right) < 0,
\end{equation}
where cylindrical coordinates $(R,\phi,z)$ are used, is violated. Constant fields or fields $B_{\phi}\propto R$, like those produced by $z$-dependent 
differential rotation from an axial field, are candidates for this instability. The most unstable Fourier mode for such a configuration is the 
nonaxisymmetric mode $m=1$ despite the smoothing action of the differential rotation. In a mean-field approach, due to the relation
\begin{equation}\label{alpha}
\alpha \simeq - \frac{\tau_{\rm corr}}{3}{\cal H},
\end{equation}
helicity ${\cal H}= {\cal H}_{\rm kin}- {\cal H}_{\rm curr}$ often indicates an existing $\alpha$-effect. ${\cal H}_{\rm kin}$ and ${\cal H}_{\rm curr}$ are the 
kinematic and current helicity, 
\begin{equation}\label{heli}
{\cal H}_{\rm kin} =\langle\vec{u}' \cdot {\rm rot}\, \vec{u}'\rangle, \quad {\cal H}_{\rm curr}= \frac{1}{\mu_0 \rho} \langle \vec{B}'\cdot {\rm rot}\vec{B}'\rangle,
\end{equation}
respectively and $\tau_{\rm corr}$ is the correlation time of the field pattern. Nonvanishing  helicity could indeed be an indication of the 
existence of a resulting  $\alpha$-effect on the basis of the TI. In case of the presence of an external field $\f{B}^0$ and with a large magnetic 
Reynolds number, a dominating contribution from magnetic-field fluctuations to the $\alpha$-effect can 
arise (Pouquet, Frisch \& Leorat 1976; Brandenburg \& Subramanian 2007). \\ 
Keeping in mind the pseudoscalar nature of helicity, a possible relation between helicity and both the external field and differential rotation would be 
\begin{equation}\label{h1}
{\cal H} \propto B^0_i\, B^0_j\, \Om_{i,j}, 
\end{equation}
where $B^0_i$ means the external field components and $\Om_{i,j}$ the gradient tensor of the basic rotation. The sign of ${\cal H}$ in Eq. (\ref{h1}) 
does not depend on the sign of the magnetic field, but it depends on the sign of the shear $\Om_{i,j}$. Consequently, the dynamo number 
\begin{equation}
D = \frac{L^2\, \alpha\, \Om}{\eta_{\rm T}^2},
\end{equation}
would now have a positive-definite sign. Here $L$ is the characteristic scale of the cylinder domain and $\eta_{\rm T}\simeq \tau_{\rm corr} \langle{u'}^2\rangle$. 
The sign of the dynamo number determines important properties of possible dynamo processes. E.g., a simple disk dynamo oscillates for a positive dynamo number and 
it is stationary for a negative dynamo number. Also, the magnetic Reynolds number of the $\alpha$-effect itself is important, that is, with Eq. (\ref{alpha}) 
including the magnetic contribution,
\begin{equation}
C_\alpha= \frac{L |\cal{H}|}{\langle u'^2\rangle}
\label{Caa}
\end{equation}
($\eta_{\rm T}$ does not include magnetic contributions, Vainshtein \& Kitchatinov 1983). We shall see that after the onset of TI, this quantity is of the
order of unity when the averages are taken as an integral over the azimuthal direction.

In the following, the details of the model used are explained. Section~\ref{sec_instability} describes the onset of the TI in a cylindrical enclosure and
the resulting field structure. In Sect.~\ref{sec_helicity}, it is shown that -- if the axial field is strong enough -- the 
numerical simulations are in good agreement with relation \ref{h1} in the case of $z$-dependent differential rotation, where it simplifies to 
${\cal H} \propto {B^0_z}^2 \mathrm{d}\Om/\mathrm{d}z$ and only a (strong enough) $z$-component $B^0_z$ of the external field is needed. In 
Sect.~\ref{sec_alpha} we show that the generated helicity is connected with an $\alpha$-effect, which is also proportional to the gradient of the 
differential rotation.

\section{Model}\label{model}
\begin{figure}[htb]
\begin{center}
\resizebox{0.7\hsize}{!}{\includegraphics[width=\textwidth]{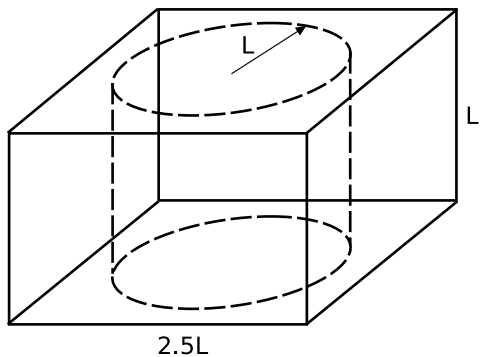}}
\caption{The considered domain is a cylinder inside a box with Cartesian grid of side length $2.5L$ and height $L$. The full set of equations is solved 
         only in the cylindrical domain.}
\label{fig_sketch_cyl}
\end{center}
\end{figure}

We consider a differentially-rotating cylinder with radius $L$ and height $L$ embedded in a box with cartesian grid
of side length $2.5L$ and height $L$ (see Fig.~\ref{fig_sketch_cyl}). The cylinder radius is 20\% less than the box side length, which appeared to be
a good compromise between wasted computing power in the corner regions and decreasing influence from the geometry of the box. We calculated
using the PENCIL code \cite{brandenb_2002}, a high-order finite difference code (sixth order in space and third order in time). Inside the cylinder 
domain, we solve both momentum and induction equations. Outside the cylinder, we kept the initial velocity to 
sustain the differential rotation via this no-slip conditions on the rim of the cylinder. On the top and the bottom, we apply stress-free conditions 
for the flow.
For the magnetic field, perfect conductor boundary conditions are applied on all box boundaries. Additionally,
in the region outside the cylinder, magnetic diffusivity $\eta$ is enhanced to a value ten times larger than inside the cylinder.\\
To keep the system nearly incompressible with a code for compressible media, the value of $\Om(z)$ is chosen in a way that the resulting maximum of the 
(meridional) velocity never exceeds 16\% of the speed of sound, which was set to $c_{\rm s}=8$. The initial density is set to $\varrho_0=1$, fluctuations 
are of the order of $10^{-4}$. The spatial resolution is $N=96\times$96$\times$48. Test runs with $N=128\times$128$\times$96 deliver only slight differences 
with smoother fields.\\
The initial magnetic field consists of the external time-independent homogeneous field $\f{B}^0=(0,0,B^0_z)$ applied only inside the cylinder. It is 
twisted into a strong toroidal field by the differential rotation. 
The strength of the generated field depends on $B^0_z$, varied between $B^0_z=0.01$ and $B^0_z=0.1$, and the gradient of the differential rotation 
$\mathrm{d}\Om/\mathrm{d}z$, where the latter was fixed to $|\mathrm{d}\Om/\mathrm{d}z| = 1$ for all presented calculations. Based on the radius 
(set to $L = 1$) and the velocity at $z=1$ ($U=\Om L=1$), and the viscosity ($\nu=0.03$), the Reynolds number has a value of $\mathrm{Re}=33$. In the following text, 
system rotation always refers to the rotation on top of the cylinder and time is given in units of rotation time. The magnetic Prandtl number $\mathrm{Pm}$ is 
varied between $\mathrm{Pm}=10$ and $\mathrm{Pm}=30$ with no qualitatively change of the instability. In the following analysis, we concentrate on $\mathrm{Pm}=15$.
We choose a large Prandtl number to encourage dynamo action, although we observed no dynamo.

\section{Instability}\label{sec_instability}
\begin{figure}[htb]
\resizebox{\hsize}{!}{\includegraphics[width=\textwidth]{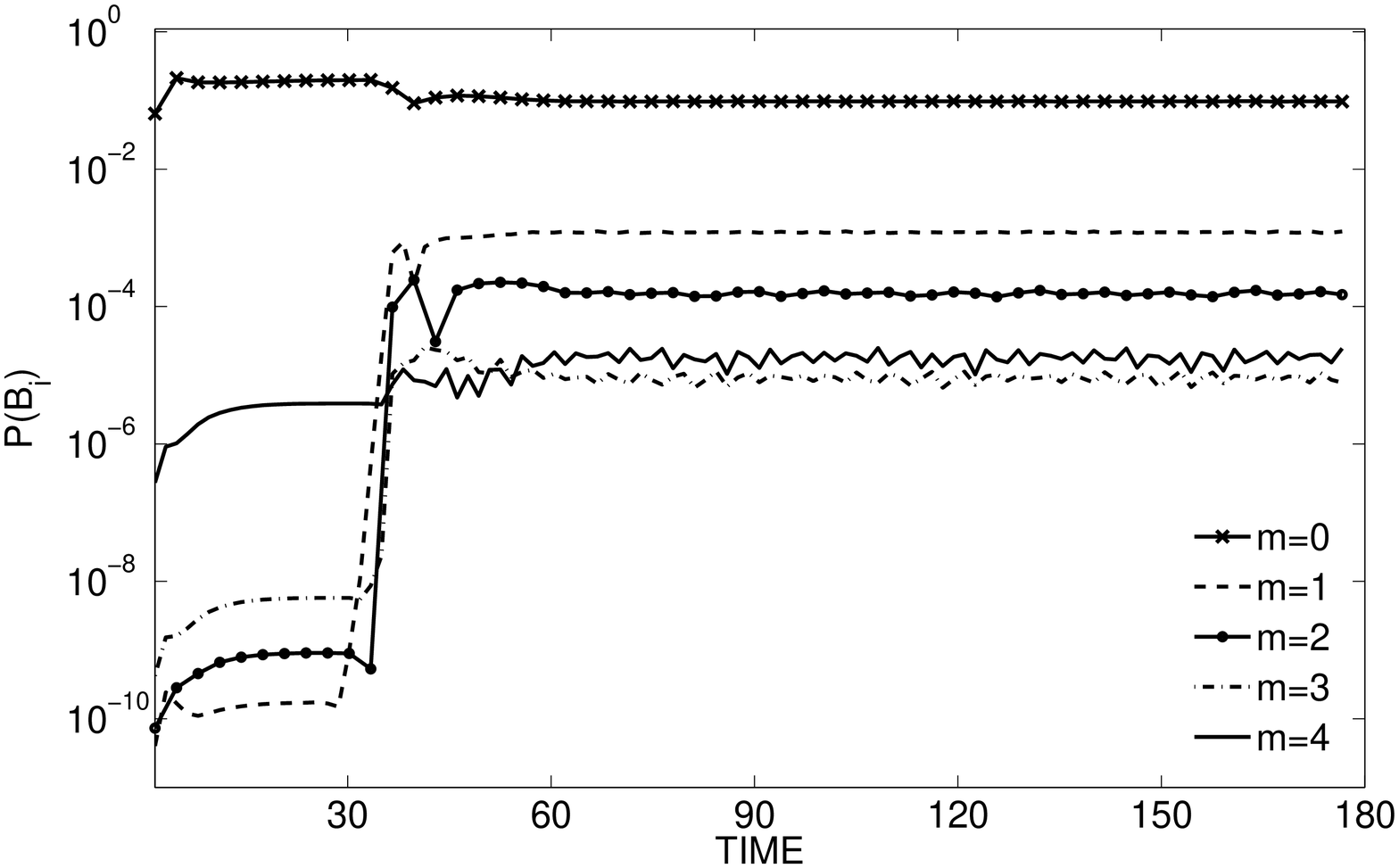}}
\caption{The power spectrum of the Fourier modes of $B_{\phi}$ for $\mathrm{Pm}=15$ and $B^0_z=0.05$. Onset of the instability 
         near $t=28$ where the dominating nonaxisymmetric mode $m=1$ starts to grow. Mode $m=4$ is nearly unaffected by the instability.}
\label{fig_bpower}
\end{figure}

If $B_{\phi}$ becomes strong enough, the TI occurs and leads to a growing nonaxisymmetric field. The largest nonaxisymmetric
mode is $m=1$. By ``strong enough" we mean that not only does the magnitude of $B_{\phi}$ reach high enough values, but also a certain 
threshold $B_{\phi}/B^0_z$ (with $B_{\phi}/B^0_z$ of the order of $\mathrm{Pm}$) needs to be crossed, which is different from the case of pure toroidal 
fields (Tayler 1973; see R\"udiger et al. 2007). This means that an additional poloidal field component suppresses the instability. For  
$\mathrm{Pm}=15$, the instability sets in at a Hartmann number of $\mathrm{Ha}=B^{\ast} R/\sqrt{\mu_0 \varrho \nu \eta}=130$, where $B^{\ast}\approx 1$ 
means the maximal value of the generated toroidal field $B_{\phi}$. To produce such a strong toroidal component an external field $B^0_z \ge 0.04$ is needed. 
The instability did not occur for $B^0_z > 0.08$.\\
The nonaxisymmetric structures appear first near the axis in the lower part of the cylinder where velocity 
magnitude is small, and grows to the steady state (Fig.~\ref{fig_bpower}). For $B_R$ and $B_z$, the $m=1$ mode becomes the largest one, in $B_{\phi}$ 
the axisymmetric mode remains dominant due to the permanently reproduced axisymmetric field. In the nonlinear regime, before the steady state is reached,
higher modes also appear. Nonetheless, their magnitudes stay below 10\% of the
$m=1$ mode. One exception is the $m=4$ mode, caused by the box geometry. During the initial phase, this mode is already present and influences the magnetic 
field outside the cylinder near the box boundaries. However, the field inside the cylinder is much less affected, and the $m=4$ mode does not seem to influence 
the nature and onset of the instability at all. Additionally, during the growth phase of all other nonaxisymmetric modes, the $m=4$ remains nearly unchanged and 
does not play an extraordinary role in the final state. Modes higher than $m=4$ are not plotted in Fig.~\ref{fig_bpower}.

\begin{figure}[htb]
\resizebox{\hsize}{!}{\includegraphics[width=\textwidth]{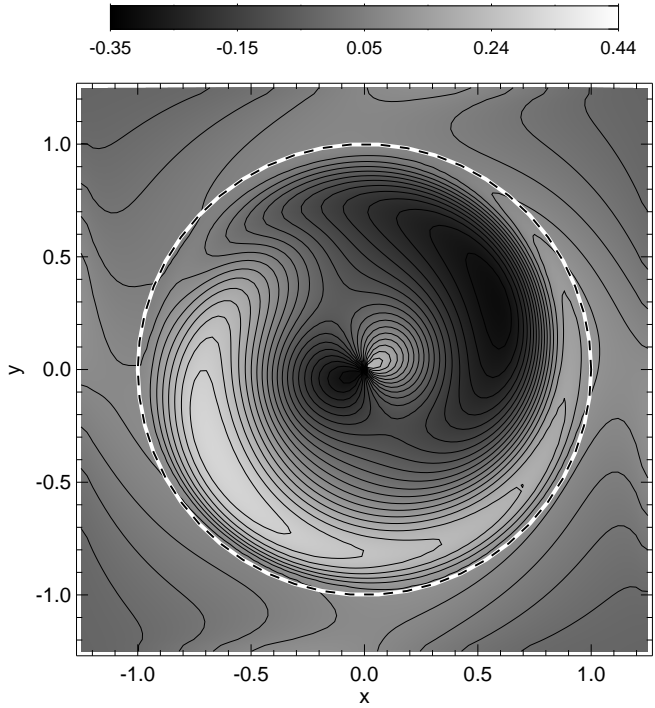}}
\caption{The $m=1$ pattern in the magnetic field $B_{\phi}$ for $\mathrm{Pm}=15$ and $B^0_z=0.05$. The pattern drifts relative to the systems rotation with
$4.5$ times the rotation period of the cylinder. The cylinder boundary is marked by the dashed black-white line.}
\label{fig_m1}
\end{figure}

The pattern of the $m=1$ mode (Fig.~\ref{fig_m1}) possesses an azimuthal drift velocity relative to the system rotation. It rotates with $4.5$ rotation 
periods of the cylinder, independent of the strength of the externally applied field. The TI in a Taylor-Couette system exhibits the same 
characteristics of the unstable mode \cite{rued_destab}. 
\begin{figure}
\resizebox{\hsize}{!}{\includegraphics[width=\textwidth]{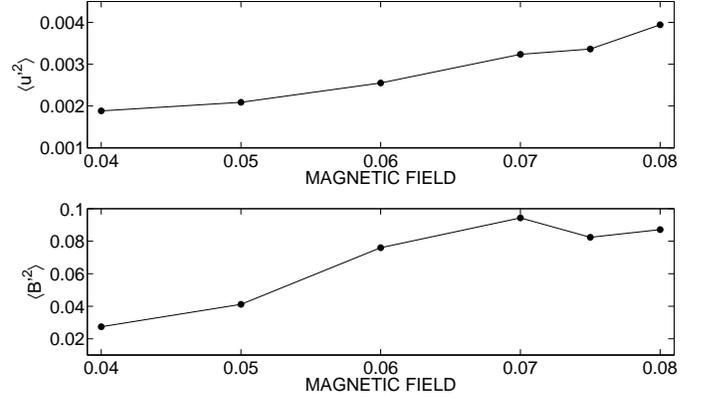}}
\caption{The external field dependence on the energy of velocity and magnetic field fluctuations at $R=0.5$ averaged in vertical
         direction for $0.35<z<0.65$.}
\label{fig_marc4}
\end{figure}
%
\section{Generation of  helicity}\label{sec_helicity}
\begin{figure}[htb]
\resizebox{\hsize}{!}{\includegraphics[width=\textwidth]{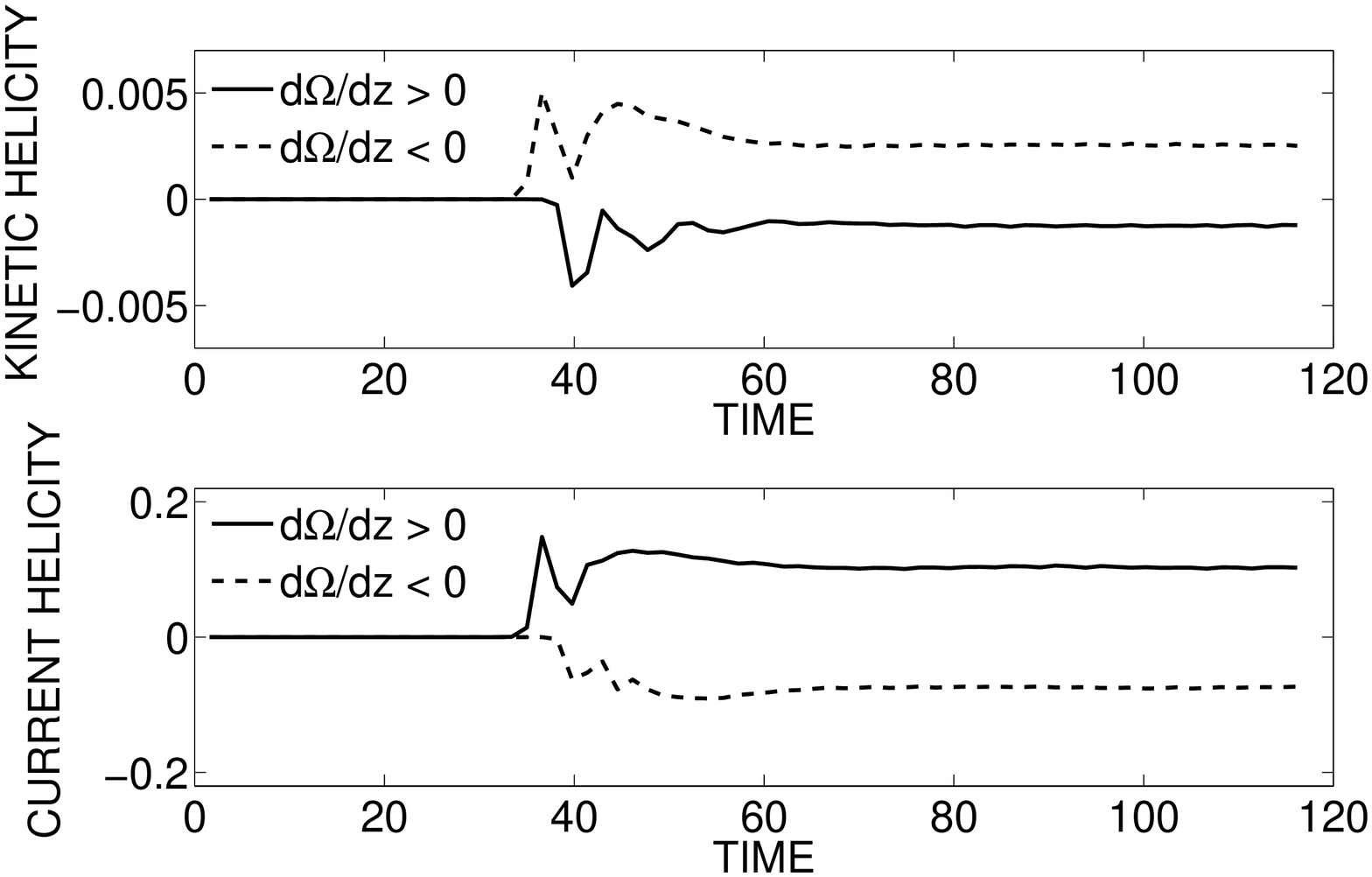}}
\caption{Kinematic helicity ${\cal H}_{\rm kin}$ and current helicity ${\cal H}_{\rm curr}$ at $R=0.5$ and $z=0.5$ for $\mathrm{Pm}=15$: both
         have different signs whereas signs of current helicity and gradient of angular velocity $\mathrm{d}\Omega/\mathrm{d}z$ are the same. The
         external field values are $B^0_z=0.05$ for negative and $B^0_z=0.06$ for positive gradient. Note the different size of 
         ${\cal H}_{\rm kin}$ and ${\cal H}_{\rm curr}$.}
\label{fig_hk_hc}
\end{figure}

\begin{figure}[htb]
\resizebox{\hsize}{!}{\includegraphics[width=\textwidth]{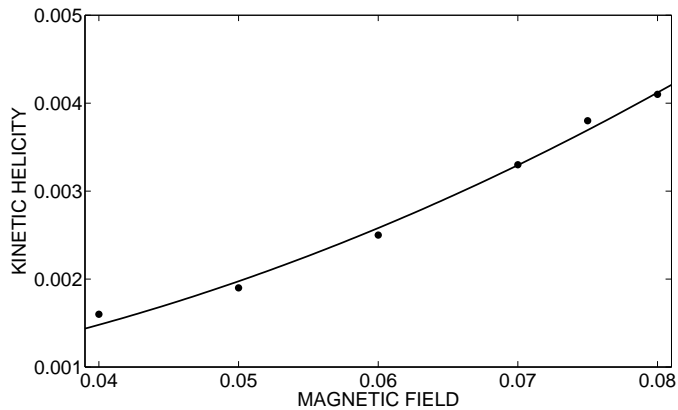}}
\caption{Kinematic helicity ${\cal H}_{\rm kin}$ as function of the external field $B^0_z$ with fitted parabola. Values of ${\cal H}_{\rm kin}$ are taken
         at $R=0.5$ and averaged along vertical direction for $0.35<z<0.65$.}
\label{fig_hk_b02}
\end{figure}

To test the relation described by Eq. (\ref{h1}) in a simple setup, we restricted differential rotation to depend only on the $z$-direction. 
\footnote{The resulting non-conservative centrifugal force drives a meridional flow that, however, remains small (see Fig.~\ref{fig_bpower}).}
For this case, the helicity ${\cal H}$ should depend on the sign of $\mathrm{d}\Om/\mathrm{d}z$, but not on the sign of the external field and 
scale with the squared value of the latter, i.e. ${\cal H} \propto {B^0_z}^2\, \mathrm{d}\Om/\mathrm{d}z$.
During the onset of the instability, the deviation of the flow structure from the original toroidal field is comparable
to that of the magnetic field. Also, here the mode $m=1$ is the largest nonaxisymmetric mode and the higher modes appear with less energy.\\
For the definition of fluctuations, we average velocity and magnetic field along the azimuthal direction, i.e.
\begin{equation}
 \langle \f{u}\rangle = \frac{1}{2\pi} \oint \f{u}\, \mathrm{d}\phi, \quad \langle \f{B}\rangle = \frac{1}{2\pi} \oint \f{B}\, \mathrm{d}\phi,
\end{equation}
and regard the nonaxisymmetric parts as fluctuations of the fields. For constant $\mathrm{d}\Om/\mathrm{d}z$,
and with a homogeneous external field, the fluctuations should be constant in the vertical direction. Due to the finite height of the cylinder, boundary effects
near the top and bottom induce additional flow disturbances and changes in the structure of the magnetic field in these regions. The independence is roughly 
preserved in the central region between $z=0.35$ and $z=0.65$. In Fig.~\ref{fig_marc4} we show the energies of the fluctuations of velocity and 
magnetic field averaged over this central region, and their dependence on the external field $B^0_z$. The values are taken in radial direction 
at $R=0.5$. The magnetic energy always exceeds the kinetic energy. The ratio varies between $15$ and $30$. Thus, the instability is dominated by
the magnetic field. This is also reflected in the helicities. We again use azimuthal-averaged quantities,
\begin{equation}
{\cal H}_{\rm kin}\!=\! \frac{1}{2\pi}\! \oint \f{u}' \cdot {\rm rot}\,\f{u}' \, \mathrm{d}\phi, \quad 
 {\cal H}_{\rm curr}\! =\! \frac{1}{2\pi\mu_0\rho}\! \oint \f{B}' \cdot {\rm rot}\,\f{B}'\, \mathrm{d}\phi.
\end{equation}
Both are zero as long as no instability exists, with proceeding instability and growing nonaxisymmetric modes the helicity 
quantities also become unequal to zero. An example plot of ${\cal H}_{\rm kin}$ for $R=0.5$ and $z=0.5$ is shown in Fig.~\ref{fig_hk_hc}.
After a transition, a steady-state value is reached for both ${\cal H}_{\rm kin}$ and ${\cal H}_{\rm curr}$. The current helicity
is approximately 40 times larger than the kinetic helicity in the example shown in the center of the cylinder. In the vertical direction we observed
the same behavior as for the fluctuations of flow and magnetic field. Near the top and bottom, the helicities are stronger than those in the region 
between $z=0.35$ and $z=0.65$. For a positive gradient in angular velocity ${\cal H}_{\rm kin}$ is negative and ${\cal H}_{\rm curr}$ positive. With the 
changing sign of $\mathrm{d}\Omega/\mathrm{d}z$, the sign of ${\cal H}_{\rm kin}$ and ${\cal H}_{\rm curr}$ also changes, whereas the sign of $B^0_z$ has no
influence. And, as expected, the absolute value of 
${\cal H}_{\rm kin}$ is proportional to the squared value of the external field. For $\mathrm{Pm}=15$ the relation between $B^0_z$ and 
${\cal H}_{\rm kin}$ is shown in Fig.~\ref{fig_hk_b02}. The fitted parabola does not exactly cross the origin. This shift is attributed to
the instability requiring a certain value of $B^0_z$ (i.e. $B^0_z=0.04$ for $\mathrm{Pm}=15$) to set in with a certain value for ${\cal H}_{\rm kin}$ larger 
than zero.
%
\section{Alpha effect}\label{sec_alpha}
\begin{figure}[htb]
\resizebox{\hsize}{!}{\includegraphics[width=\textwidth]{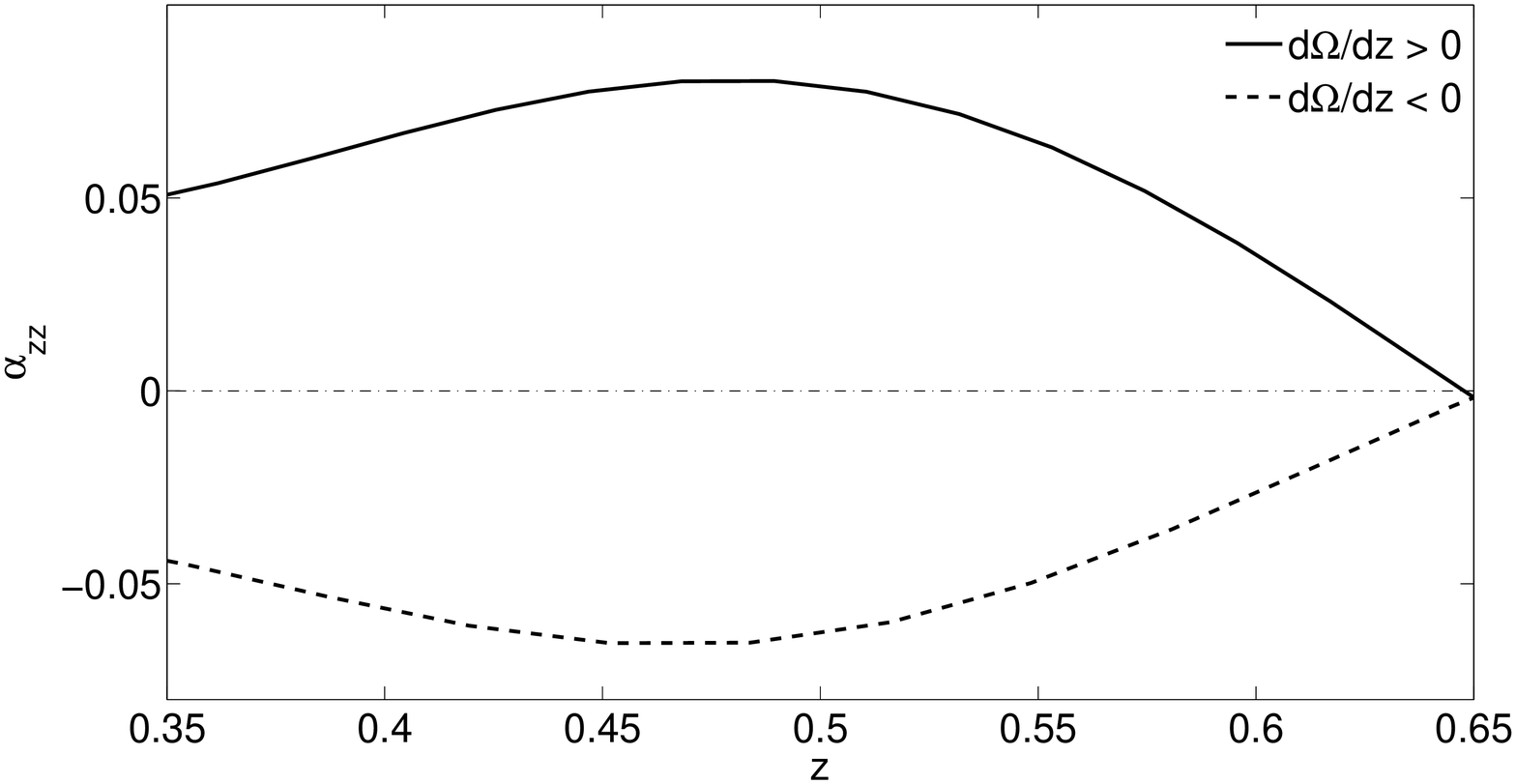}}
\caption{Coefficient of the $\alpha$-effect $\alpha_{zz}$ at $R=0.5$ for positive and negative $\mathrm{d}\Om/\mathrm{d}z$. Sign of $\alpha_{zz}$ 
         is the same as ${\cal H}_{\rm curr}$ and $\mathrm{d}\Om/\mathrm{d}z$.}
\label{fig_alpha}
\end{figure}

Equation (\ref{alpha}) suggests that the changing sign of the helicities by changing the direction of $\mathrm{d}\Om/\mathrm{d}z$ will also
appear for the term $\alpha_{zz}$ in the $\alpha$-tensor responsible for the regeneration of the magnetic field. Indeed, calculating
$\alpha_{zz}={ \langle\f{u}' \times \f{B}'\rangle}_z/B^0_z$ using the $z$-component of the electromotive force, gives values for 
$\alpha_{zz}$ with the same sign as $\mathrm{d}\Om/\mathrm{d}z$ for $z<0.65$ (see Fig.~\ref{fig_alpha}). 
Near the top boundary, the disturbances already discussed for the helicities change the values of $\alpha_{zz}$ dramatically to the opposite sign. We 
excluded this region and took mean values of $\alpha_{zz}$ in the same range $0.35<z<0.65$ used for the helicities and find coefficients between 
$\alpha_{zz} = 0.038$ for $B^0_z=0.04$ and $\alpha_{zz} = 0.067$ for $B^0_z=0.08$. The correlation time derived from expression (\ref{alpha}) using
both helicities, $\tau_{\rm corr} = 3 \alpha_{zz} / ({\cal H}_{\rm curr} - {\cal H}_{\rm kin})$, is of the order of the system rotation time. Note that 
the $\alpha$-effect here is mainly due to the current helicity and the product $\alpha_{zz}{\cal H}_{\rm curr}$ is 
positive.
If the external field is switched off, all magnetic field modes decay in the investigated parameter region of low Reynolds numbers and magnetic Prandtl numbers
of the order of ten. The  helicities drop down to zero nearly immediately. Despite the high magnetic Prandtl number, we do not observe any dynamo action, 
as found in comparable geometry \cite{braithw_2006}. Also, the reported cyclic behavior of the instability did not occur in our simulations.
%
\section{Conclusions}
We have demonstrated how the instability of toroidal magnetic fields leads to helicity generation in the nonaxisymmetric parts of a flow without density 
stratification. The suggested relation between helicity, external field, and differential rotation, ${\cal H} \propto B^0_i\, B^0_j\, \Om_{i,j}$, 
is rather well fulfilled in the case of $z$-dependent differential rotation. Both the kinematic helicity and current helicity depend on the squared 
value of the $z$-component of the external magnetic field $B^0_z$, and scale linearly with the gradient of the differential rotation. Also, the magnetically
dominated $\alpha$-effect depends on the direction of $\mathrm{d}\Om/\mathrm{d}z$, that is $\alpha_{zz}$ holds the same sign as $\mathrm{d}\Om/\mathrm{d}z$.\\
The realized model is too simple to estimate consequences of an $\alpha$-effect based on this new kind of helicity production for environments without 
density stratification and possibly for new dynamo models. In a next step, therefore, we would like to check the relation (\ref{h1}) when differential rotation depending 
on the distance from the rotation axis is present as well. With an appropriate flow it becomes 
${\cal H}_{\rm kin} \propto B^0_R B^0_z \mathrm{d}\Om/\mathrm{d}R$. In this case, helicity generation should be observable only if both components 
of the external field are unequal to zero. The problem with this constellation is the changing sign of the product of both field components in one hemisphere
in the simplest configuration, a dipole-like field.

\end{document}